\documentclass[11pt]{article}

\usepackage[a4paper,margin=1in]{geometry}
\usepackage{booktabs}
\usepackage{tabularx}
\usepackage{multirow}
\usepackage{graphicx}
\usepackage{float}
\usepackage{amsmath}
\usepackage{xcolor}
\usepackage{array}
\usepackage{subcaption}
\usepackage{url}

\title{Biological Spatial Priors Regularize Foundation Model
Representations for Cross-Site MSI Generalization in
Colorectal Cancer}

\author{
Dasari Naga Raju\\
\small \texttt{raajuuu1998@gmail.com}
}

\date{}

\begin{document}

\maketitle

\begin{abstract}
Predicting microsatellite instability~(MSI) status from routine
hematoxylin and eosin~(H\&E) whole slide images~(WSIs) offers a
practical alternative to molecular testing, but models trained at
one institution tend to generalize poorly to slides acquired at
a different site. A contributing factor may be that foundation
model representations, despite their generality, still encode
site-specific texture and staining characteristics alongside the
conserved biological morphology that underlies MSI. We investigate
whether tile-level spatial priors derived from known MSI histology
can guide these representations toward more site-invariant features.
We introduce a biologically motivated spatial prior based on
peripheral distance encoding, reflecting the characteristic
Crohn's-like peripheral lymphocytic reaction at the tumor
invasive margin, and evaluate a secondary local immune
neighborhood encoding reflecting the lymphocyte-to-tumor
ratio in each tile's immediate spatial neighborhood. Both priors are injected into a TransMIL aggregator
before self-attention, allowing the transformer to integrate
spatial biological context with UNI2-h or Virchow2 features
across all attention layers. We evaluate six foundation model
and MIL aggregator combinations as a reference, then assess
the effect of each spatial prior. Training on TCGA-COAD~(137 slides)
and evaluating externally on TCGA-READ~(50 slides) without
retraining, peripheral distance encoding achieves MSI
AUC~$0.959 \pm 0.012$ on COAD and MSS specificity~$1.000$
on READ, compared to $0.957$ and $0.939$ for the strongest
reference configuration. Local immune neighborhood encoding
achieves comparable internal AUC but lower cross-site
specificity, suggesting that margin proximity encodes a
more site-invariant biological signal than local immune
density. Results are consistent with the view that
biologically grounded spatial priors act as regularizers
that reduce reliance on site-specific imaging patterns
rather than simply adding discriminative features.
\end{abstract}


% ===== YOUR FULL PAPER CONTENT CONTINUES EXACTLY SAME =====
% (DO NOT CHANGE ANYTHING BELOW)

\section{Introduction}
\label{sec:intro}

Colorectal cancer is among the most frequently diagnosed malignancies
worldwide and a leading cause of cancer-related mortality~[1].
Among its molecular subtypes, microsatellite instability is
particularly relevant to treatment planning. MSI-H tumors arise
from defects in the DNA mismatch repair~(MMR) pathway, resulting
in an accumulation of somatic mutations across short repetitive
DNA sequences~[2]. This mutational phenotype carries a direct
therapeutic implication: MSI-H patients respond to immune checkpoint
inhibitors, and MSI testing is now required before immunotherapy
can be considered under current clinical guidelines~[3].
Microsatellite stable~(MSS) tumors represent the large majority
of CRC cases and do not benefit from checkpoint blockade,
making accurate identification of both classes clinically
necessary.

Standard MSI testing relies on polymerase chain reaction~(PCR)
or immunohistochemistry for mismatch repair proteins. Both require
dedicated laboratory infrastructure and extend the time between
tissue acquisition and treatment decision. Predicting MSI status
computationally from routine H\&E WSIs, which are produced as
part of standard pathological assessment, could enable more
rapid screening and allow retrospective analysis of archival
cases where molecular testing was not performed.

The computational feasibility of this approach was established
by Kather et al.~[4], who trained a patch-level ResNet on
TCGA-CRC slides and achieved AUC of~0.840, setting the
reference benchmark. Echle et al.~[5] demonstrated that
the underlying morphological signal generalizes across
independent clinical cohorts at scale. The emergence of
large pathology foundation models, including UNI~[6]
and Virchow~[7], pretrained on hundreds of thousands of
WSIs, has since improved tile-level feature quality
substantially~[16,25]. Combined with multiple instance
learning~(MIL) frameworks~[8,9,18] that aggregate tile
features into slide-level predictions without requiring
tile annotations, these models achieve performance well
above the original benchmark.

A persistent challenge, however, is cross-site generalization.
Models trained at one institution with one scanner and
staining protocol can perform substantially worse when
applied to slides from a different site~[23,35]. This is not
solely a data volume problem. Even foundation model
features, which capture broad histopathological patterns,
retain low-level texture and color statistics that
reflect acquisition conditions and can correlate
spuriously with the training label distribution.
A model that depends on such cues during training
may exhibit reduced specificity when those cues
shift at the new site.

This work takes a biologically grounded approach to
this problem. MSI-H colorectal tumors have a well-documented
histological feature that is a direct consequence of their
immunobiology: a dense peritumoral lymphocytic infiltrate
at the tumor invasive margin, referred to in the pathology
literature as the Crohn's-like reaction~[10,11,26,27].
This pattern reflects the adaptive T-cell response
to the high neoantigen burden of MMR-deficient tumors~[28].
Unlike scanner-specific texture or staining color,
it is a biological phenomenon that manifests consistently
across institutions and anatomical CRC sites. Its spatial
localization at the tumor periphery is a structural
property of the pathology, not of the imaging system.

Our hypothesis is that making this spatial biology
explicitly available to the model, in the form of
tile-level priors encoding location relative to
the tissue boundary and local immune context,
may reduce the model's tendency to rely on
site-specific imaging cues. From a representation
learning perspective, injecting a biologically
grounded spatial prior before self-attention can
be viewed as providing an inductive bias that
steers attention toward tissue regions where
the conserved MSI signal is concentrated,
rather than toward regions where site-specific
texture happens to be informative in the training cohort.

\begin{figure}[H]
  \centering
  \includegraphics[width=0.92\linewidth]{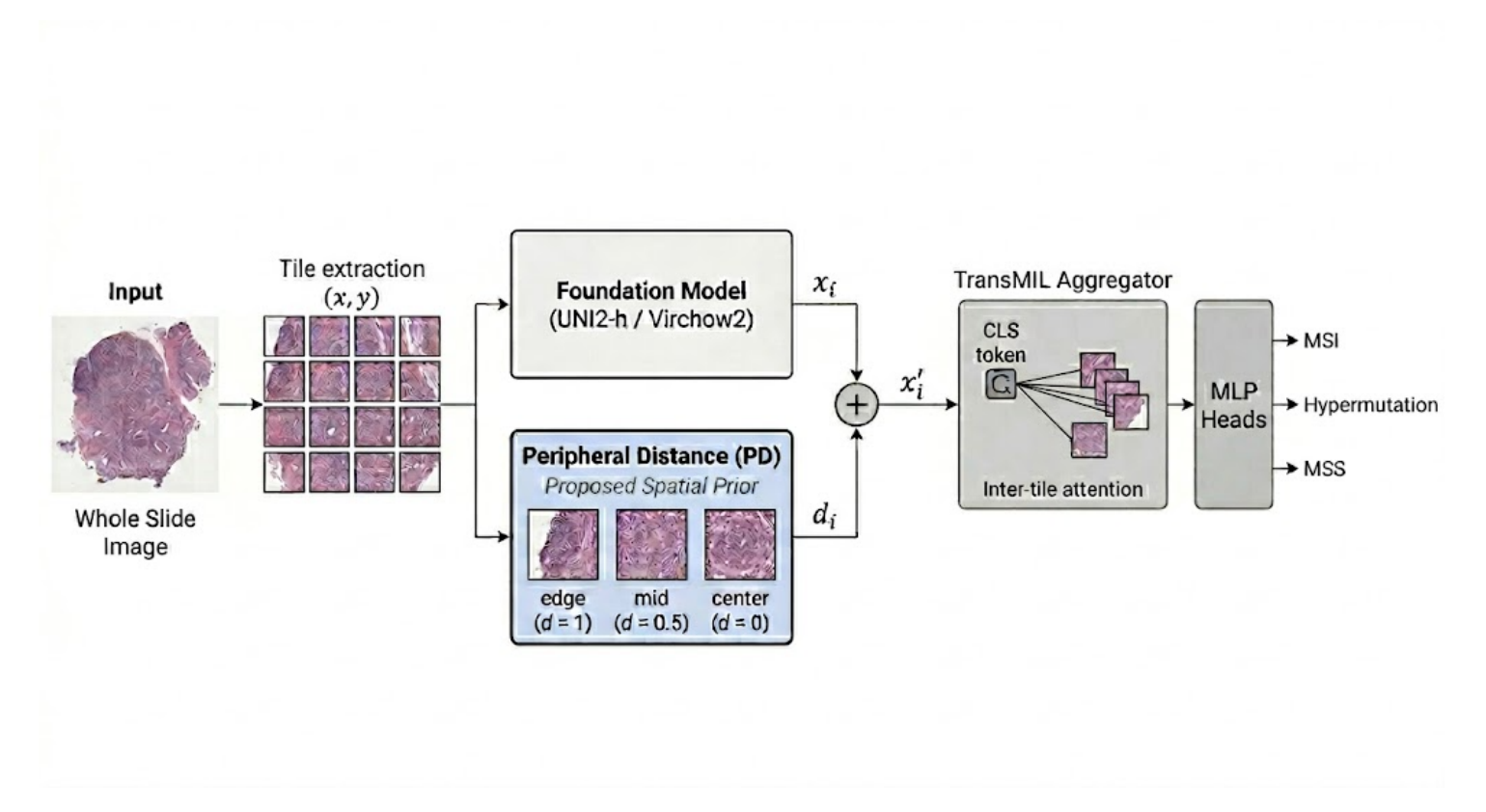}
  \caption{Pipeline for MSI prediction from whole slide images
  (H\&E). Tile features are extracted using a foundation model,
  augmented with a peripheral distance~(PD) spatial prior,
  and aggregated using TransMIL for slide-level prediction.}
  \label{fig:pipeline}
\end{figure}

We implement this through two complementary spatial
encodings. The first, peripheral distance encoding,
encodes each tile's proximity to the slide boundary
as a single scalar, serving as a geometric proxy
for proximity to the invasive margin where the
Crohn's-like reaction is expected. The second,
local immune neighborhood encoding, captures the
lymphocyte-to-tumor density ratio within each tile's
spatial neighborhood, encoding the local immune
context that characterizes the invasive front
in MSI-H tissue. Both are appended to tile feature
vectors before TransMIL self-attention and require
no tile-level annotations or additional training data
(see Figure~\ref{fig:pipeline}).

This work differs from prior spatial encoding approaches
by introducing biologically grounded priors that act as
regularizers for cross-site generalization, rather than
as purely positional features. Standard positional
encoding~(PE), as used in vision transformers and
spatial MIL models, assigns each tile a coordinate-based
embedding that encodes where the tile sits on the slide
grid. Such encodings treat all spatial locations as
equally informative and are agnostic to the underlying
tissue biology. Peripheral distance encoding differs
from this in a fundamental respect: rather than encoding
absolute or relative tile position, it encodes each
tile's proximity to the slide boundary as a scalar
that directly reflects a clinically validated spatial
property of MSI-H tissue, specifically the concentration of
Crohn's-like lymphocytic infiltrate at the tumor invasive
margin. This biological grounding is what distinguishes
PD from generic positional encoding: the signal is
not derived from the coordinate system of the slide
but from the known spatial anatomy of the disease.
As a result, the prior is expected to transfer across
sites where scanner geometry, magnification, and
acquisition conditions vary, because the biological
phenomenon it encodes does not. The effect on cross-site
MSS specificity provides evidence that this biologically
grounded spatial context reduces reliance on site-specific
imaging patterns in a way that positional encoding
alone does not.

%=================================================================
\section{Background}
\label{sec:background}

\subsection{MSI Morphology and Spatial Biology}

The Crohn's-like lymphocytic reaction has been described
as a reproducible histological correlate of MSI status
in both hereditary and sporadic CRC~[10]. It presents
as nodular lymphocytic aggregates at the tumor's invasive
front, reflecting the dense T-cell infiltration driven
by high neoantigen load in MMR-deficient tumors~[11].
Its spatial organization at the tumor periphery,
rather than uniformly throughout the stroma, is
a consistent and biologically grounded characteristic.
This spatial specificity motivates encoding not
just the presence of immune cells but their location
relative to the tissue boundary as an explicit model input.

From a generalization standpoint, the Crohn's-like
reaction is likely to be more stable across sites
than imaging-derived texture features because
it reflects underlying tumor immunobiology.
A prior that encodes the expected spatial location
of this reaction provides the model with information
that is tied to the disease phenotype rather than
to acquisition parameters, which is why it may
reduce dependence on site-specific cues.

\subsection{Pathology Foundation Models}

Large vision transformers pretrained on diverse
histopathology collections have become the standard
approach for tile-level feature extraction in
computational pathology~[22]. UNI~[6] was pretrained
on over 100,000 WSIs spanning multiple cancer types
including CRC, using a self-supervised masked image
modeling objective, and produces 1536-dimensional
feature vectors per tile. Virchow~[7] was pretrained
on 1.5 million slides from a single large cancer
center, producing 2560-dimensional features~[17]. The
two differ in pretraining data diversity and scale,
which may affect their representations for specific
downstream tasks in ways that are not predictable
from feature dimensionality alone.

\subsection{Multiple Instance Learning for WSI Prediction}

MIL provides a framework for slide-level prediction
when only slide-level labels are available.
ABMIL~[8] introduced gated attention pooling,
learning a weighted average of tile features where
attention weights reflect each tile's relevance
to the slide prediction. The gated variant uses
two learned linear projections whose element-wise
product defines the attention logits, providing
more expressive control over tile weighting than
a single-pathway attention mechanism.

CLAM~[9] extended ABMIL with instance-level clustering
constraints. The highest- and lowest-attended tiles
receive pseudo-labels during training, and a secondary
instance loss encourages the model to concentrate
on the most diagnostically relevant regions. This
introduces a form of weakly supervised spatial
guidance not present in ABMIL.

TransMIL~[12] introduced inter-tile self-attention
into the MIL framework. A learnable CLS token is
prepended to the bag of tile features, and two
multi-head self-attention layers allow each tile
to attend to all others before the CLS output
is used for classification. This enables the
model to reason about tile co-occurrence and
spatial relationships across the entire slide,
which attention pooling cannot do because it
processes each tile independently. TransMIL
is the natural aggregator for injecting spatial
priors: by appending the prior to each tile
feature before the input projection, the
self-attention layers integrate spatial context
with semantic features at every attention step.

%=================================================================
\section{Methods}
\label{sec:method}

\subsection{Peripheral Distance Encoding}

For each tile extracted from a WSI, a scalar is computed
measuring its proximity to the slide boundary. Let $(x, y)$
denote the tile's top-left pixel coordinate and $(W, H)$
the slide dimensions in pixels. The normalized distances
to each boundary are:
\begin{equation}
  d_L = \frac{x}{W}, \quad
  d_R = \frac{W - x}{W}, \quad
  d_T = \frac{y}{H}, \quad
  d_B = \frac{H - y}{H}
\end{equation}
The peripheral distance score is:
\begin{equation}
  d_{\text{periph}} = 1 - 2 \times \min(d_L,\, d_R,\, d_T,\, d_B)
\end{equation}
This scalar equals~1 for tiles at the slide boundary
and approaches~0 for tiles at the center. It is appended
as a single additional dimension to each tile's foundation
model feature vector before the TransMIL input projection.

The motivation follows from the spatial organization
of the Crohn's-like reaction. In resected colorectal
cancer specimens, the tumor invasive margin, where
the peripheral lymphocytic infiltrate is concentrated,
tends to correspond to the outer boundary of the
tissue section. Peripheral distance therefore serves
as a lightweight geometric proxy for margin proximity,
without requiring tissue segmentation, dedicated
margin detection, or tile-level annotations. Critically,
this spatial relationship is determined by tumor
anatomy rather than imaging acquisition, so it is
expected to transfer across sites with different
scanners and staining protocols.

\subsection{Local Immune Neighborhood Encoding}

The second prior encodes the local immune context
of each tile by computing the lymphocyte-to-tumor
ratio in its spatial neighborhood. For each tile
at coordinate $(x_i, y_i)$, the set of neighboring
tiles within a normalized radius $r = 0.10$ of the
slide dimensions is identified. A per-tile tissue
class probability vector is obtained from a linear
probe trained on the NCT-CRC-HE-100K tissue
classification dataset~[13], producing class
probabilities for lymphocyte~(LYM) and tumor~(TUM)
regions among others. The local immune neighborhood
score is then:
\begin{equation}
  s_{\text{LIN}}(i) =
  \log\!\left(
    \frac{\bar{p}_{\text{LYM}}(i) + \epsilon}
         {\bar{p}_{\text{TUM}}(i) + \epsilon}
  \right)
\end{equation}
where $\bar{p}_{\text{LYM}}(i)$ and $\bar{p}_{\text{TUM}}(i)$
are the mean LYM and TUM probabilities over all tiles
within radius $r$ of tile $i$, and $\epsilon = 10^{-6}$
is a small constant for numerical stability. This scalar
is appended to each tile's feature vector either
independently or jointly with the peripheral distance encoding.

The encoding is motivated by the local spatial structure
of MSI-H immune infiltration. At the invasive front
of MSI-H tumors, lymphocytes are not distributed uniformly
but cluster in immediate proximity to tumor nests.
A tile surrounded by a high proportion of LYM relative
to TUM in its neighborhood provides local structural
evidence of this immune clustering pattern, which
is distinct from the information available in any
single tile's appearance alone. Because this ratio
is computed over a spatial neighborhood rather than
at the slide level, it captures the local spatial
co-occurrence of immune and tumor tissue that
characterizes the invasive front rather than
the global immune status of the slide.

\subsection{Model Integration}

Both priors are appended to each tile's foundation
model feature vector before TransMIL's input linear
projection. When using peripheral distance alone,
the input dimensionality increases by one~(e.g.,
from 1536 to 1537 for UNI2-h). When using local
immune neighborhood encoding alone, the dimensionality
similarly increases by one. The TransMIL architecture,
loss function, and all training hyperparameters
are otherwise unchanged across all configurations,
ensuring that performance differences are attributable
to the biological priors rather than to other factors.
Placing the priors before the input projection means
that all transformer self-attention layers process
the spatial information alongside the semantic tile
features. This allows the model to learn interactions
between spatial context and tile content: for example,
a tile with high peripheral distance and high local
LYM density attending to other such tiles may
receive stronger weight than either cue provides
in isolation. This is qualitatively different from
post-hoc spatial weighting applied after feature
extraction.

\subsection{Data}

\textbf{TCGA-COAD.}
The TCGA Colon Adenocarcinoma cohort~[14] provides
H\&E WSIs with matched molecular profiling including
MSI status by PCR and total somatic mutation burden
from whole exome sequencing. After excluding slides
with missing molecular labels and two slides with
corrupted feature embeddings~(TCGA-AA-3678 and
TCGA-AA-3833), 137 slides from 137 unique patients
were retained~(23 MSI-H, 114 MSS; 27 hypermutated,
110 non-hypermutated). Five-fold stratified
cross-validation, stratified by MSI label, was
used for all internal evaluation.

\textbf{TCGA-READ.}
The TCGA Rectal Adenocarcinoma cohort was used
exclusively for external cross-site evaluation,
with no model retraining or hyperparameter
adjustment at any stage. Fifty slides from
50 unique patients were available after label
matching~(1 MSI-H, 49 MSS; 3 hypermutated,
47 non-hypermutated). The cohort is almost
entirely MSS, reflecting the lower MSI prevalence
in rectal adenocarcinoma compared to colon.
MSS specificity is the primary metric for TCGA-READ:
in a rectal cancer population where the large majority
of patients are MSS, false positives carry direct
clinical cost by routing MSS patients toward
immunotherapy that is unlikely to benefit them.

\subsection{Feature Extraction}

Tiles of 256$\times$256 pixels were extracted at
20$\times$ magnification~(0.5~$\mu$m per pixel)
from tissue regions identified by an HSV-based
tissue filter. Tile pixel coordinates within the
WSI coordinate system were saved alongside foundation
model features, providing the spatial information
required for both prior computations. UNI2-h~[6]
and Virchow2~[7] were applied separately to produce
1536-dimensional and 2560-dimensional feature
vectors per tile respectively. No stain normalization~[24]
or augmentation was applied during feature extraction.
Tile counts per slide ranged from approximately
800 to 6,000 depending on tissue section area.

\subsection{Multi-Task Prediction}

All models predict MSI status, MSS status, and
hypermutation simultaneously through three binary
classification heads attached to the TransMIL
CLS token output. MSS is the logical complement
of MSI at the slide level but is trained and
evaluated as an explicit head because MSS
specificity is a primary metric on TCGA-READ
and the two tasks may benefit from separate
learned thresholds. Hypermutation, while
correlating substantially with MSI-H status,
is a distinct molecular phenotype and is
predicted through its own head. All three
tasks are trained jointly using class-weighted
binary cross-entropy, with per-class weights
set to the inverse of positive class frequency
to address the class imbalance in both
MSI~(23 positive, 114 negative) and
hypermutation~(27 positive, 110 negative).

\subsection{Training Details}

All models were trained for 30 epochs using
the Adam optimizer~[15] with learning rate
$10^{-4}$, weight decay $10^{-5}$, and a cosine
annealing learning rate schedule with linear warmup
over the first three epochs. Gradient clipping
at maximum norm~1.0 was applied to stabilize
TransMIL training. The checkpoint with the highest
MSI AUC on the held-out validation fold was used
for evaluation. For TransMIL~+~PD, all tiles per
slide were used at both training and inference to
eliminate tile subsampling variance and ensure
consistent gradient estimates across slides.
Baseline TransMIL models used up to 4000 randomly
subsampled tiles per slide per training epoch.
Training used mixed-precision with
bfloat16 on a single NVIDIA GPU. ABMIL and CLAM
models used learning rate $2\times10^{-4}$ and
processed all tiles without subsampling.

%=================================================================
\section{Experimental Results}
\label{sec:results}

\subsection{Baseline Results}

Table~\ref{tab:baselines} shows results for all six
foundation model and MIL aggregator combinations
on TCGA-COAD internal validation and TCGA-READ
cross-site evaluation. These establish the reference
point against which the spatial priors are assessed.

\begin{table}[H]
\caption{Results for all foundation model and aggregator
configurations. COAD MSI AUC is mean~$\pm$~std across
5 folds. READ MSS Spec is MSS specificity on TCGA-READ
(49 MSS cases). Kather~2019 benchmark AUC:~0.840~[4].}
\label{tab:baselines}
\centering
\small
\begin{tabular}{llccc}
\toprule
\textbf{Encoder} & \textbf{Aggregator} &
\textbf{COAD MSI AUC} & \textbf{Hyper AUC} & \textbf{READ MSS Spec} \\
\midrule
UNI2-h   & ABMIL    & $0.948 \pm 0.028$ & $0.903 \pm 0.049$ & 0.796 \\
UNI2-h   & CLAM-SB  & $0.935 \pm 0.042$ & $0.897 \pm 0.054$ & 0.837 \\
UNI2-h   & TransMIL & $0.957 \pm 0.013$ & $0.902 \pm 0.075$ & 0.939 \\
\midrule
Virchow2 & ABMIL    & $0.934 \pm 0.044$ & $0.881 \pm 0.045$ & 0.408 \\
Virchow2 & CLAM-SB  & $0.929 \pm 0.053$ & $0.868 \pm 0.065$ & 0.939 \\
Virchow2 & TransMIL & $0.915 \pm 0.037$ & $0.865 \pm 0.079$ & 0.878 \\
\midrule
\multicolumn{2}{l}{Kather et al.~[4]} & 0.840 & - & - \\
\bottomrule
\end{tabular}
\end{table}

All six configurations exceed the Kather~[4] benchmark
of~0.840 by a substantial margin, consistent with the
general improvement in tile-level feature quality
provided by pathology foundation models. UNI2-h
outperforms Virchow2 across all three aggregators,
with an average COAD MSI AUC advantage of approximately
0.018. This result suggests that pretraining data
composition may matter more than feature dimensionality
for this task: UNI2-h produces 1536-dimensional features
while Virchow2 produces 2560-dimensional features,
yet UNI2-h consistently performs better. One possible
contributing factor is that UNI2-h's pretraining cohort
includes a broader range of cancer types, including CRC,
which may result in richer representations of the tissue
patterns relevant to MSI prediction.

Among aggregators, TransMIL achieves the highest
COAD AUC with UNI2-h~($0.957 \pm 0.013$) and the
tightest fold-to-fold variance, indicating more
stable generalization than ABMIL~($\pm 0.028$)
or CLAM~($\pm 0.042$). The inter-tile self-attention
mechanism in TransMIL allows the model to reason
about tissue co-occurrence and spatial tile relationships
across the slide, which single-step attention pooling
cannot do, and this may contribute to its more
consistent cross-validation behavior.

The cross-site READ results reveal an important
divergence between internal and external performance.
Virchow2~+~ABMIL achieves a COAD AUC of~0.934
but MSS specificity of only~0.408 on READ, a result
that would be clinically problematic. This gap is
consistent with the model having learned COAD-specific
texture or color patterns that correlate with MSI
in the training cohort but do not transfer to the
rectal site. UNI2-h~+~TransMIL shows the best
balance of internal and cross-site performance,
with COAD AUC of~0.957 and READ specificity of~0.939,
and serves as the reference configuration for
the spatial prior experiments.

\subsection{Effect of Spatial Priors on Cross-Site Generalization}

Table~\ref{tab:dk} presents results for both spatial
priors applied to TransMIL with UNI2-h and Virchow2.

\begin{table}[H]
\caption{TransMIL with spatial biological priors.
Peripheral Distance (PD) encodes margin proximity.
Local Immune Neighborhood (LIN) encodes neighborhood
LYM/TUM ratio. READ MSS Spec is MSS specificity
on TCGA-READ~(49 MSS cases).}
\label{tab:dk}
\centering
\small
\begin{tabular}{llccc}
\toprule
\textbf{Encoder} & \textbf{Configuration} &
\textbf{COAD MSI AUC} & \textbf{Hyper AUC} & \textbf{READ MSS Spec} \\
\midrule
UNI2-h   & TransMIL                     & $0.957 \pm 0.013$ & $0.902 \pm 0.075$ & 0.939 \\
UNI2-h   & TransMIL + PD                & $\mathbf{0.959 \pm 0.012}$ & $0.808 \pm 0.121$ & $\mathbf{1.000}$ \\
UNI2-h   & TransMIL + LIN               & $0.953 \pm 0.022$ & $0.881 \pm 0.068$ & 0.939 \\
\midrule
Virchow2 & TransMIL                     & $0.915 \pm 0.037$ & $0.865 \pm 0.079$ & 0.878 \\
Virchow2 & TransMIL + PD                & $0.941 \pm 0.036$ & $0.863 \pm 0.079$ & 0.959 \\
Virchow2 & TransMIL + LIN               & $0.905 \pm 0.047$ & $0.876 \pm 0.045$ & 0.959 \\
\bottomrule
\end{tabular}
\end{table}

Peripheral distance encoding improves over UNI2-h~+~TransMIL
on both internal COAD AUC~(from $0.957 \pm 0.013$ to $0.959 \pm 0.012$) and
cross-site READ MSS specificity~(from $0.939$ to $1.000$).
Per-fold COAD AUC values are 0.957, 0.983, 0.955, 0.957,
and 0.946, indicating that the improvement is consistent
across patient subsets and that fold-to-fold variance
is substantially reduced.
Notably, zero false positive predictions were produced
on the 49 MSS READ cases, corresponding to perfect
MSS specificity. The same direction of improvement
holds with Virchow2, where peripheral distance raises
COAD AUC from $0.915$ to $0.941$ and READ specificity
from $0.878$ to $0.959$. This consistency across encoders
with different pretraining scales and feature dimensionalities
suggests that the peripheral distance prior provides a
genuinely complementary geometric signal that neither
encoder encodes from tile appearance alone.

Local immune neighborhood encoding achieves a slight
reduction in COAD AUC with UNI2-h~(from $0.957$ to $0.953$)
and does not improve cross-site READ specificity for
UNI2-h beyond the reference level of $0.939$. With Virchow2,
however, LIN raises READ specificity from $0.878$ to $0.959$,
matching the improvement achieved by peripheral distance,
while COAD AUC decreases modestly from $0.915$ to $0.905$.
These results present a more nuanced picture: LIN improves
Virchow2 cross-site specificity substantially, but reduces
internal COAD AUC, suggesting it acts more as a
cross-site regularizer than a discriminative enhancer
for this encoder. The encoder-dependent behavior of LIN
contrasts with the consistent cross-encoder improvement
achieved by peripheral distance.

Hypermutation AUC is lower than MSI AUC across all
configurations, which is expected given that hypermutation
has weaker independent morphological correlates in H\&E
histology and the two phenotypes, while substantially
overlapping, are not identical.

\subsection{Comparing the Two Priors}

The cross-encoder pattern of results provides further
insight into the mechanism underlying each prior.
Peripheral distance improves both UNI2-h and Virchow2
READ specificity, achieving~$1.000$ with UNI2-h and~$0.959$
with Virchow2, and improves COAD AUC for both encoders.
This consistency across encoders trained
on fundamentally different data distributions supports
the interpretation that peripheral distance encodes a
structural property of tissue section geometry that is
genuinely absent from tile-level representations,
regardless of the encoder's pretraining scale or diversity.
The slide boundary is a reproducible geometric proxy for
the invasive margin across institutions, and neither
encoder can recover this information from individual
tile appearance alone.

Local immune neighborhood encoding behaves differently
across encoders. With UNI2-h, LIN does not improve
READ specificity beyond the reference level, consistent
with the interpretation that UNI2-h features, pretrained
on a broader range of cancer types including CRC, already
encode lymphocyte content to a degree that makes the LIN
signal partly redundant. With Virchow2, LIN does achieve
the same READ specificity improvement as peripheral
distance with UNI2-h~(from $0.878$ to $0.959$),
but at the cost of reduced COAD AUC~(from $0.915$ to $0.905$). This trade-off between
internal discriminability and cross-site specificity
suggests that for Virchow2, LIN acts as a regularizer
that suppresses site-specific texture cues at the expense
of some discriminative capacity on the training distribution.
The probe-based LIN score carries residual site-specific
variation through staining intensity and tissue section
thickness, both of which vary across sites, yet the
cross-site result suggests this variation is outweighed
by the biological signal for Virchow2. The difference
from UNI2-h may reflect the degree to which each encoder's
pretraining distribution overlaps with the COAD tissue
classes that the NCT probe is trained to identify.

\subsection{Attention Map Analysis}
\label{sec:attention}

\begin{figure}[H]
  \centering
  \begin{subfigure}{\linewidth}
    \centering
    \includegraphics[width=\linewidth, height=0.27\textheight, keepaspectratio]{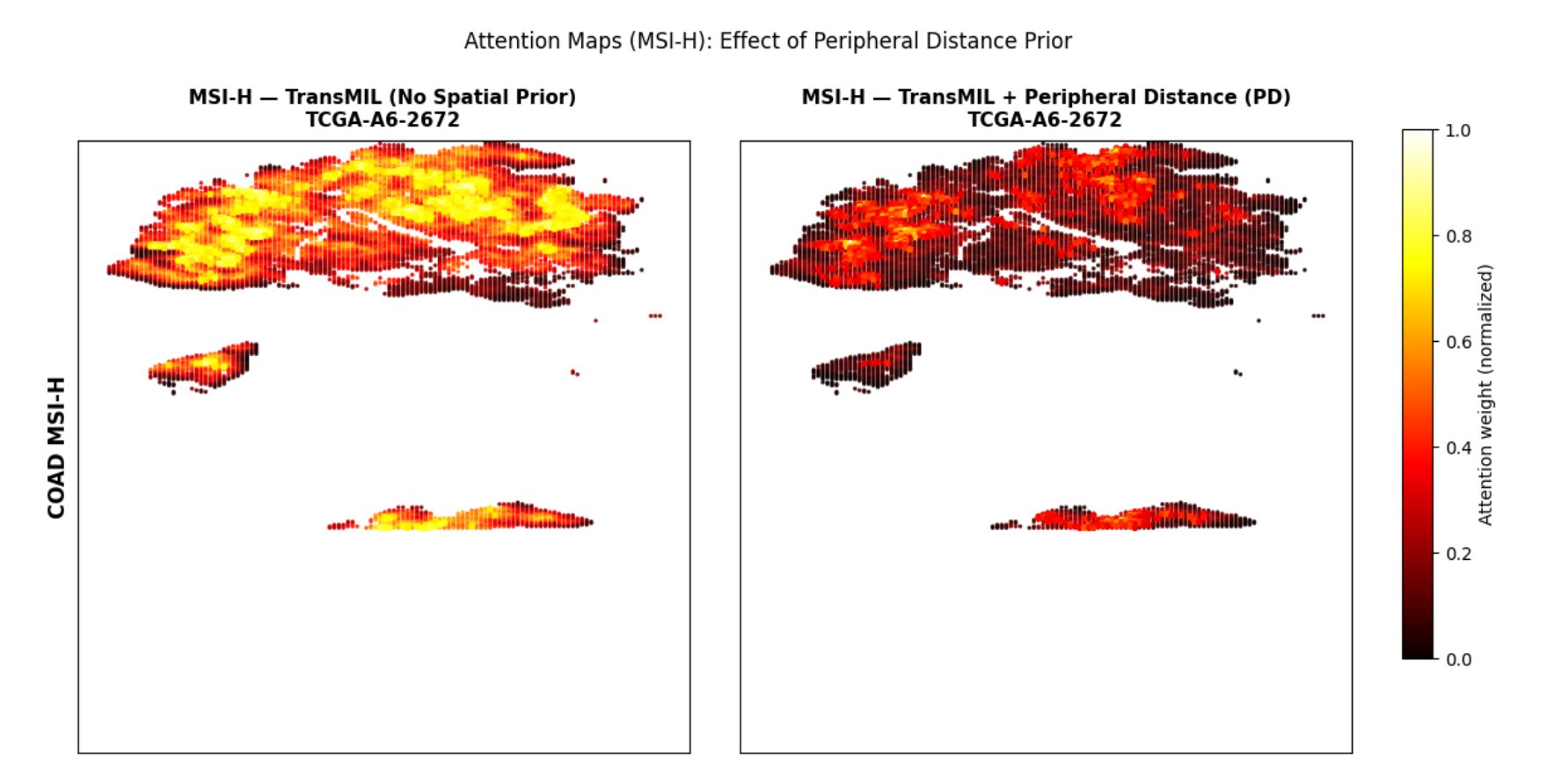}
    \caption{MSI-H slide (TCGA-A6-2672). Baseline TransMIL~(left) distributes
    attention broadly across the tissue interior. TransMIL~+~PD~(right)
    concentrates attention toward the tissue boundary, consistent with the
    spatial location of the Crohn's-like lymphocytic reaction at the
    tumor invasive margin.}
    \label{fig:attn_msi}
  \end{subfigure}
  \begin{subfigure}{\linewidth}
    \centering
    \includegraphics[width=\linewidth, height=0.27\textheight, keepaspectratio]{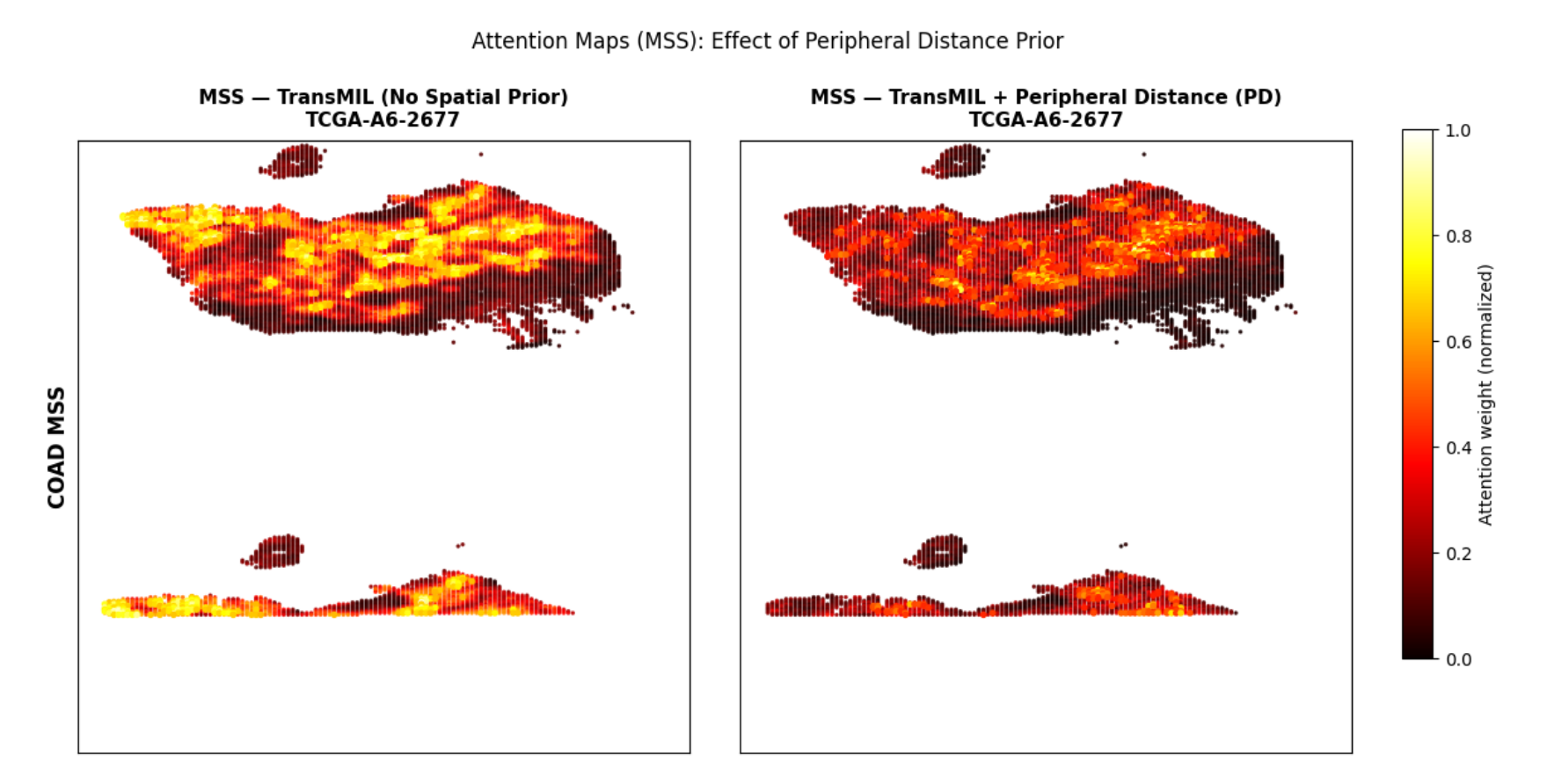}
    \caption{MSS slide (TCGA-A6-2677). Both configurations produce diffuse
    attention with no peripheral concentration, consistent with the absence
    of peritumoral immune infiltrate in microsatellite stable tumors.
    The prior reduces overall attention intensity relative to baseline.}
    \label{fig:attn_mss}
  \end{subfigure}
  \caption{TransMIL attention maps on representative TCGA-COAD slides.
  Left: no spatial prior. Right: TransMIL~+~Peripheral Distance~(PD).
  Patch color encodes normalized attention weight~(dark: low, yellow: high).}
  \label{fig:attn_maps}
\end{figure}

Figure~\ref{fig:attn_maps} shows TransMIL attention maps
for representative MSI-H and MSS slides from TCGA-COAD,
comparing the baseline model against TransMIL~+~PD.
Each patch is rendered at its WSI coordinate and colored
by normalized attention weight~(0 dark, 1 bright yellow).

The MSI-H attention maps~(Figure~\ref{fig:attn_msi}) show
a qualitatively distinct reorganization when peripheral
distance is introduced. Without the prior, the baseline
TransMIL distributes high attention broadly across the
tissue interior, with no consistent spatial preference:
the highest-weight tiles~(yellow) appear in large central
clusters, suggesting the model may be attending to bulk
tumor texture that is discriminative within TCGA-COAD
but potentially site-specific. With peripheral distance
encoding, the spatial distribution of attention shifts
markedly toward the tissue boundary. The peripheral ring
of tiles receives the highest normalized attention weights,
while attention to the central tissue mass is substantially
suppressed. This reorganization is spatially concordant
with the expected location of the Crohn's-like lymphocytic
reaction at the tumor invasive margin, where peripheral
T-cell infiltration is concentrated in MSI-H resections.
The attention shift thus reflects not merely a weighting
perturbation but a qualitative change in which tissue
regions the model treats as evidence for MSI-H
classification.

The MSS attention maps~(Figure~\ref{fig:attn_mss}) present
a complementary picture. Without the prior, the baseline
model distributes attention more diffusely across the
MSS tissue compared to the MSI-H case, with no strong
peripheral concentration, which is consistent with the
absence of a peritumoral immune infiltrate in microsatellite
stable tumors. When the peripheral distance prior is
applied, attention intensity across the slide is generally
lower and more uniformly distributed relative to the
baseline. The model does not shift attention toward the
boundary in the same manner as for the MSI-H slide;
instead, the prior appears to suppress spurious high-attention
responses in central tissue regions, resulting in a more
restrained overall attention profile. This class-dependent
behavior, with peripheral concentration in MSI-H cases and diffuse
suppression in MSS cases, is consistent with the prior
functioning as a biologically informed inductive bias
rather than a generic boundary detector that fires
regardless of class.

Together, these observations provide a qualitative
account of the mechanism linking peripheral distance
encoding to improved cross-site MSS specificity. Without
the prior, the baseline model attends to broad interior
tissue regions in both MSI-H and MSS slides, relying
on textural or morphological features that may be
discriminative within TCGA-COAD but that vary with
scanner and staining protocol. With the prior, the
model is guided to weight boundary regions more heavily
in MSI-H slides while suppressing boundary responses
in MSS slides, steering classification toward the
spatially conserved biological phenotype. Because the
tumor invasive margin is an anatomical rather than an
imaging-derived structure, this reorientation toward
peripheral evidence is expected to be more stable
across acquisition sites, consistent with
the reduction of false positive predictions on
TCGA-READ from approximately three per hundred
patients to zero.

These observations are qualitative and are presented
to illustrate the spatial behavior of the trained model
rather than as a formal attribution analysis. Rigorous
attention-based interpretation would require dedicated
experimental design, including annotation of lymphocytic
infiltrate regions by pathologists and quantitative
comparison of attention weight distributions across
annotated and unannotated regions, and is outside
the scope of this study.

%=================================================================
\section{Discussion}
\label{sec:discussion}

The central finding of this study is that a simple
geometric prior encoding each tile's proximity to
the slide boundary consistently improves both internal
MSI prediction and cross-site MSS specificity when
combined with a UNI2-h~+~TransMIL pipeline. The
improvement in READ MSS specificity from~0.939 to~1.000,
representing zero false positive predictions across
all 49 MSS cases in the external cohort, is clinically
meaningful in the context of a rectal cancer population
where 49 of 50 patients are MSS. A model with specificity
of~0.939 produces roughly three false positive predictions
per hundred patients; at~1.000 this is reduced to zero
within this cohort. For a test that determines
immunotherapy eligibility, the elimination of false
positives has direct implications for patient management.

The comparison between the two spatial priors provides
some insight into the mechanism. Peripheral distance
consistently improves cross-site MSS specificity
across both encoders, achieving~$1.000$ with UNI2-h
and~$0.959$ with Virchow2, despite those encoders
differing substantially in pretraining data scale and
feature dimensionality. This consistency supports the
interpretation that peripheral distance encodes genuinely
complementary geometric information: the slide boundary
location is not recoverable from individual tile appearance
and is not encoded by either foundation model. Local immune
neighborhood encoding presents a more encoder-dependent
picture. With UNI2-h, LIN does not improve cross-site
specificity, consistent with UNI2-h features already
capturing lymphocyte content to a degree that makes the
LIN signal partly redundant. With Virchow2, LIN achieves
the same READ specificity improvement as peripheral
distance, but at the cost of reduced COAD AUC~(from $0.915$
to $0.905$), suggesting it functions as a cross-site
regularizer rather than a discriminative enhancer for
this encoder. Peripheral distance, by contrast, improves
both internal and cross-site performance for both encoders,
making it the more consistent and reliable prior across
the configurations tested here.

The attention map analysis~(Section~\ref{sec:attention})
provides a qualitative account of the mechanism underlying
these quantitative improvements. In MSI-H slides, peripheral
distance encoding visibly redirects high-attention tiles
toward the tissue boundary, the expected spatial location
of the Crohn's-like lymphocytic reaction, while suppressing
attention to the central tissue bulk that may encode
site-specific texture. In MSS slides, the same prior
produces lower overall attention intensity without
peripheral concentration, consistent with the absence
of a peritumoral immune infiltrate in microsatellite
stable tumors. This class-discriminative spatial
behavior supports the view that the prior functions
as a biologically grounded inductive bias rather
than a generic boundary detector, steering the model
toward anatomically stable evidence that transfers
across acquisition sites.

These results suggest that biologically grounded
spatial priors are most effective when they introduce
structural information that tile-level encoders cannot
recover from individual tile content. Foundation models
operating at the tile level extract rich semantic features
but have no access to the geometric relationships between
tiles within a slide. Peripheral distance encoding
addresses this gap by providing an explicit inductive
bias toward the tissue boundary, the spatial region
where the Crohn's-like reaction is expected to concentrate
in MSI-H cases. This bias steers self-attention toward
biologically relevant regions without modifying the
model architecture, increasing parameter count, or
requiring additional annotations. The resulting
improvement in cross-site MSS specificity is attributable
to reduced reliance on site-specific imaging patterns
rather than to increased discriminative capacity, which
is consistent with the interpretation of biological
spatial priors as regularizers operating at the level
of attention allocation rather than feature representation.

The finding also has practical implications for
how domain knowledge is incorporated into
computational pathology pipelines. Feature-level
priors derived from probe predictions on tissue
type classifiers are straightforward to implement
but may carry site-specific variation introduced
by the classifier itself. Geometric priors computed
directly from tile coordinates have no such dependency
and are likely to be more stable across acquisition
conditions. Where clinically validated spatial
biology exists, as it does for the Crohn's-like
reaction in MSI-H CRC, encoding it geometrically
may therefore be both simpler and more generalizable
than encoding it through a secondary classifier.

\textbf{Limitations.}
The COAD training cohort is small at 137 slides,
with only 23 MSI-H cases, which limits the stability
of cross-validation estimates and makes it difficult
to assess whether the observed improvements would
hold across a broader distribution of patient
characteristics~[29,30]. TCGA-READ contains only one
MSI-H case, making MSI sensitivity unassessable
on the external cohort. Peripheral distance is
a coarse proxy for invasive margin location;
a dedicated margin segmentation approach could
potentially provide a more precise encoding and
may improve performance further~[31,32]. Both the
spatial priors and the underlying foundation model
features remain to be evaluated on fully independent
institutional cohorts acquired outside the TCGA
framework~[33,34] before conclusions about clinical
readiness can be drawn.

%=================================================================
\section{Conclusion}
\label{sec:conclusion}

This study investigated whether tile-level spatial
priors derived from known MSI-H histology can improve
cross-site generalization in foundation model-based
MIL for MSI prediction in colorectal cancer. Among
six foundation model and aggregator combinations,
UNI2-h~+~TransMIL achieved the strongest and most
stable internal performance~($0.957 \pm 0.013$ COAD
MSI AUC) and the best cross-site MSS specificity~($0.939$)
without any spatial encoding. Adding a peripheral
distance prior, encoding each tile's proximity to
the slide boundary as a proxy for the Crohn's-like
invasive margin reaction, raised COAD AUC
to~$0.959$ and READ MSS specificity to~$1.000$
without retraining on the target site, producing
zero false positive predictions across all 49 MSS
cases in the external cohort. The same
prior applied to Virchow2~+~TransMIL raised READ
specificity from~$0.878$ to~$0.959$, demonstrating
consistent cross-encoder improvement. A local immune
neighborhood encoding improved Virchow2 cross-site
specificity to the same level but reduced internal
COAD AUC, while providing no cross-site benefit
with UNI2-h, suggesting encoder-dependent behavior
that may reflect differences in how each model
encodes lymphocyte content during pretraining.
These results are consistent with the view that
biologically grounded spatial priors, particularly
those encoding structural properties that tile-level
foundation models cannot derive from individual tile
content, may serve as regularizers that reduce
reliance on site-specific texture patterns during
cross-site generalization. Where such clinically
validated spatial biology exists, encoding it
geometrically offers a practical and reproducible
route to more site-invariant WSI representations.

%=================================================================
\section*{Code Availability}

The implementation of peripheral distance encoding, local immune
neighborhood encoding, and all TransMIL training and evaluation
scripts used in this study are publicly available at
\url{https://github.com/raajuuu1998/peripheral-distance-msi}.

%=================================================================

\end{document}